\documentclass[useAMS,usenatbib]{mn2e}
\usepackage{epsfig}
\usepackage{float}

\usepackage{natbib}

\usepackage{color}

\usepackage{aas_macros}

\def\sige{\mbox{$\sigma_{\rm e}$}}

\def\Re{\mbox{$R_{\rm e}$}}

\def\ML{\mbox{$M/L$}}
\def\dimf{\mbox{$\delta_{\rm IMF}$}}
\def\Yst{\mbox{$\Upsilon_\star$}}

\def\mst{\mbox{$M_{\star}$}}

\def\Mvir{\mbox{$M_{\rm vir}$}}
\def\cvir{\mbox{$c_{\rm vir}$}}

\def\lsim{\mathrel{\rlap{\lower3.5pt\hbox{\hskip0.5pt$\sim$}}
    \raise0.5pt\hbox{$<$}}}                % less than or approx. symbol
\def\gsim{~\rlap{$>$}{\lower 1.0ex\hbox{$\sim$}}}

\def\a0{\mbox{$a_{\rm 0}$}}

\def\gN{\mbox{$g_{\rm N}$}}

\def\atlas3d{ATLAS$^{\rm 3D}$}

\title[MOND and IMF]{MOND and IMF variations in early-type galaxies from \atlas3d}

\author[Tortora C. et al.]{\noindent
C.~Tortora$^{1,5}$\thanks{E-mail: ctortora@physik.uzh.ch},
A.J.~Romanowsky$^{2,3}$, V.F.~Cardone$^{4}$,
N.R.~Napolitano$^{5}$, Ph.~Jetzer$^{1}$
\\~\\
$^1$ Universit$\ddot{a}$t Z$\ddot{u}$rich, Institut f$\ddot{u}$r
Theoretische Physik, Winterthurerstrasse 190, CH-8057,
Z$\ddot{u}$rich, Switzerland \\
$^2$ Department of Physics and Astronomy, San Jos\'e State
University, San Jose, CA 95192, USA\\
$^3$ University of
California Observatories, 1156 High
Street, Santa Cruz, CA 95064, USA \\
$^4$ INAF -- Osservatorio Astronomico di Roma, via Frascati 33,
00040 - Monte Porzio Catone, Roma, Italy\\
$^5$ INAF -- Osservatorio Astronomico di Capodimonte, Salita
Moiariello, 16, 80131 - Napoli, Italy\\}

\begin{document}
\date{Accepted  Received }
\pagerange{\pageref{firstpage}--\pageref{lastpage}} \pubyear{xxxx}
\maketitle

\label{firstpage}
\begin{abstract}
MOdified Newtonian dynamics (MOND) represents a phenomenological
alternative to dark matter (DM) for the missing mass problem in
galaxies and clusters of galaxies. We analyze the central regions
of a local sample of $\sim 220$ early-type galaxies from the
\atlas3d\ survey, to see if the data can be reproduced without
recourse to DM. We estimate dynamical masses in the MOND context
through Jeans analysis, and compare to \atlas3d\ stellar masses
from stellar population synthesis. We find that the observed
stellar mass--velocity dispersion relation is steeper than
expected assuming MOND with a fixed stellar initial mass function
(IMF) and a standard value for the acceleration parameter \a0.
Turning from the space of observables to model space, a) fixing
the IMF, a universal value for \a0\ cannot be fitted, while, b)
fixing \a0\  and leaving the IMF free to vary, we find that it is
``lighter'' (Chabrier-like) for low-dispersion galaxies, and
``heavier'' (Salpeter-like) for high dispersions. This MOND-based
trend matches inferences from Newtonian dynamics with DM, and from
detailed analysis of spectral absorption lines, adding to the
converging lines of evidence for a systematically-varying IMF.
\end{abstract}


\section{Introduction}\label{sec:intro}

Flat rotation curves in spiral galaxies (\citealt{Rubin_Ford70}),
dynamics and gravitational lensing in early-type galaxies (ETGs;
ellipticals and lenticulars), and in clusters of galaxies
(\citealt{Romanowsky+03}; \citealt{Bradac+08};
\citealt{Napolitano+09_PNS}; \citealt{Tortora+10lensing};
\citealt{Napolitano+11_PNS}) are usually modelled using the
classical Newtonian theory of gravity. In this context, vast
amounts of dark matter (DM) are inferred, in consonance with the
standard cosmology (e.g., \citealt{Hinshaw+13_WMAP9}), and with
the modern understanding of galaxy evolution as seeded by the
collapse of DM haloes (e.g., \citealt{deLucia+06}).

Unfortunately, the nature of DM is still not clear, with no direct
experimental detection of DM particles. In this context, it should
be recalled that Newtonian dynamics has never been experimentally
tested in the extremely weak-field limit as in the outskirts of
galaxies. An alternative phenomenological framework was proposed
by \cite{Milgrom83b, Milgrom83}, in which Newton's second law of
dynamics becomes $F = m g$, where the acceleration $g$ is related
to the Newtonian one \gN\ by $g \, \mu(g/\a0) = \gN$. Here,
\a0~$\sim cH_0$ is an universal constant and $\mu(x)$ is an
arbitrary function with the limiting behaviours $\mu(x\gg1) = 1$
and $\mu(x\ll1) = x$.

This model, referred to as MOdified Newtonian Dynamics (MOND),
reproduces the flat rotation curves of spiral galaxies without
recourse to undetectable DM, and provides a natural explanation
for the observed relation between galaxy rotation and luminosity
\citep{TF77,Sanders_McGaugh02} or baryonic mass
(\citealt{McGaugh+12}). Thirty years after its introduction, MOND
remains remarkably successful on galaxy scales, but the
conclusions to date have been largely based on late-type galaxies.
Only a few analyses have been carried out on ETGs (e.g.,
\citealt{Cardone+11MOND}; \citealt{Ferreras+12_MOND_TEVES};
\citealt{Milgrom+12_Xray_ell}), and it is not clear if they can be
integrated consistently into the MOND framework.

The difficulty with ETGs has been the lack of a large, homogeneous
sample with high-quality dynamical analysis.  These criteria are
not yet met for the ideal case where kinematical data extend to
large radii, but the advent of the \atlas3d\ survey
\citep{Cappellari+11_ATLAS3D_I} entails a remarkable opportunity
to test MOND in the centers of ETGs.

\atlas3d\ provides a sample of 260 local ETGs with central masses
estimated both by dynamics and by stellar population synthesis
(SPS). The latter aspect is critical since the stars comprise the
dominant component of the central mass, even in models with DM
included. However, standard SPS modelling is hindered by the
uncertain stellar initial mass function (IMF), and the \atlas3d\
team have taken a purely dynamical approach, where the total mass
is decomposed into stars and DM, assuming Newtonian gravity and
standard DM halo models. The resulting stellar masses imply strong
variations in the IMF, in agreement with many recent studies
(\citealt{Conroy_vanDokkum12b}; \citealt{Cappellari+12,
Cappellari+13_ATLAS3D_XX}; \citealt{Spiniello+12};
\citealt{Dutton+13}; \citealt{Ferreras+13};
\citealt{Goudfrooij_Kruijssen13};
\citealt{LaBarbera+13_SPIDERVIII_IMF}; \citealt{TRN13_SPIDER_IMF};
\citealt{Weidner+13_giant_ell}).

Our aim in this paper is to revisit the \atlas3d\ results in the
context of MOND.  Can the central dynamics of ETGs be reproduced
with MOND and a standard, fixed IMF?  Alternatively, is MOND
consistent with current claims for a variable IMF? The dynamical
approach we adopt provides an estimate for the  IMF
"normalization", which we cannot unambiguously relate to the slope
of the bottom- or top-end of the IMF. Throughout the present paper
and in agreement with other works we will interpret our results in
terms of variations in the fraction of low-mass stars.

The paper is organized as follows. In Sect. \ref{sec:data} we
describe our dynamical methods and the data to be analyzed. In
Sect.  \ref{sec:results} we discuss the results of the paper,
which are the constraints on the acceleration scale and on the
IMF. Conclusions are made in Sect. \ref{sec:conclusions}.

\section{Methods}\label{sec:data}

We perform our analysis on a sample of local ETGs from the
\atlas3d\ survey (\citealt{Cappellari+13_ATLAS3D_XV,
Cappellari+13_ATLAS3D_XX}). About 15\% of the full sample have
significant gradients of the stellar mass-to-light ratio (\ML)
implied by their young stellar populations (H$\beta$ equivalent
width greater than 2.3~\AA), so we omit these cases and retain a
sample of 224 galaxies.

The relevant data for each galaxy include a) the effective radius,
\Re, b) the projected stellar velocity dispersion, \sige, within a
circularized aperture of radius \Re, the $r$-band c) total
luminosity $L_r$ and d) stellar \ML\ ($\Upsilon_*$) derived by SPS
fitting of the spectra with \cite{Vazdekis+12} models and a
\cite{Salpeter55} IMF. The \cite{Chabrier01} IMF yields stellar
masses that are $\sim 0.26$ dex smaller.

It is important to note that the published $L_r$ and \Re\ values
are not self-consistent. The former correspond to detailed
multi-gaussian expansion (MGE) fits that extend to typically
$\sim$~4~\Re. The latter are the MGE-based values renormalized by
a factor of 1.35 to correspond to more conventional estimates from
the literature.  Here we will use these \Re\ values, but adjust
each $L_r$ value such that the projected luminosity inside \Re\
for our adopted de Vaucouleurs model is the same as in the
original MGE model.  This extrapolation means $L_r$ is typically
increased by a factor of $\sim 1.2$.

The basic assumptions of MOND are as follows.
\begin{enumerate}
\item Standard dynamics is not valid in the limit of low
accelerations, such that the gravitational acceleration $g(r)$
differs from the Newtonian one $\gN(r) = G M_{\rm tot}/r^{2}$,
where $M_{\rm tot}$ is the total mass involved (DM + stars). The
MONDian $g(r)$ reduces to the Newtonian one at high accelerations.
\item In the low-acceleration limit, the acceleration is given
by $(g/\a0)g = \gN$, where \a0 is the MOND acceleration constant.
Thanks to this limit the rotation curves are flat and it is
possible to recover the \cite{TF77} relation.
\item The transition from the Newtonian regime to the low
acceleration regime occurs around a characteristic acceleration
scale \a0\ (\citealt{Milgrom83b}). Unless otherwise stated, we
adopt the standard value of $\a0 = 1.2\times10^{-10}$~m~s$^{-2}$,
as calibrated from spiral galaxy dynamics (\citealt{Begeman+91}).
\end{enumerate}

To connect the low- and high-acceleration regimes, a general
formula is needed, which reduces to the low-acceleration limit as
in (ii). The following expression is adopted:
\begin{equation}
g(r) \mu \left[ \frac{g(r)}{\a0} \right] = \gN(r) ,
\label{eq:MOND}
\end{equation}
where $\mu(x)$ is an empirical ``interploating'' function, with
the properties $\mu(x\gg1) = 1$ and $\mu(x\ll1) = x$. One recovers
the Newtonian theory when $\mu(x)=1$ and the deep MOND regime when
$\mu(x)=x$. We adopt the following expressions: a) our reference
choice $\mu_{1}(x)= x/(1+x)$ (\citealt{Famaey_Binney05};
\citealt{Angus08}) and b) $\mu_{2}(x)= x/\sqrt{1+x^{2}}$, which
was the first one successfully tested
(\citealt{Sanders_McGaugh02}).

Our dynamical approach is based on the spherical Jeans equations,
relating the acceleration $g$ to the mass as follows:
\begin{equation}
\frac{d[j(r) \sigma_{r}^{2}(r)]}{dr} +\frac{2 \beta(r)}{r} j(r)
\sigma_{r}^{2}(r)=-\rho(r) g (r), \label{eq:jeans}
\end{equation}
where $j(r)$ is the deprojected luminosity profile, $\sigma_{r}$
is the radial velocity dispersion and $\beta(r) = 1- \sigma_{\rm
\theta}^{2}/\sigma_{\rm r}^{2}$ is the velocity dispersion
anisotropy (e.g., \citealt{Sanders00,Cardone+11MOND}). We adopt
isotropic models (i.e. $\beta(r)=0$) as our default, but we will
also examine the impact of anisotropy.

We assume no DM, thus $M_{\rm tot} = \mst$ (from SPS) and $\gN(r)
= G \mst(r)/r^{2}$. We approximate the deprojected de Vaucouleurs
profile with an analytic expression from \cite{PS96}. Assuming
that \Yst\ is constant with radius, the mass density profile is
$\rho(r) = \Yst \, j(r)$ and the mass profile $\mst(r)$ is easily
derived (see \citealt{Cardone+11MOND}). Thus, in
Eq.~(\ref{eq:jeans}), $j(r)$, $\mu(x)$ and $\beta(r)$ and $g(r)$
are given and $\sigma_r$ can be derived by simple integration.
Finally, to match the observed aperture averaged velocity
dispersion \sige, we project $\sigma_r$ along the line of sight
and within a circular aperture (see \citealt{ML05a, ML05b};
\citealt{Tortora+09}).

\section{Results}\label{sec:results}

\subsection{Faber--Jackson relation}

\begin{figure}
\centering \psfig{file=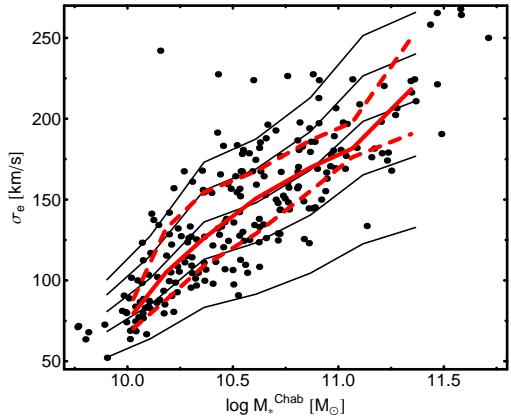,
width=0.4\textwidth}\caption{Stellar mass vs.\ velocity dispersion
for \atlas3d\ galaxies, where a default Chabrier IMF is assumed.
The black solid lines show the predictions from Jeans equations
adopting the median \Re--\mst\ relation from the observations,
\a0\ fixed to the standard value and different IMFs, corresponding
to  $\dimf =0.5,1,1.5,2,2.5$ from bottom to top. The black points
are the observations, the red solid and dashed lines are the
median and 25-75th quantiles. The observed trend is slightly
steeper than the MOND prediction for any fixed IMF.}\label{fig:FJ}
\end{figure}

We begin with vanilla MOND-modelling assumptions of fixed \a0\ and
IMF, and investigate how well a critical observable is reproduced:
the relation between stellar mass and velocity dispersion
(\citealt{FJ76}). We compare the \atlas3d\ observations with
predictions from Jeans equations, where we have adopted the median
\Re--\mst\ relation from the observations, fixed \a0\ to the
standard value and assumed a varying IMF, parameterized in terms
of the ``IMF mismatch parameter'' $\dimf\ \equiv
\Yst/\Upsilon_{\star,{\rm MW}}$. The latter relates the dynamical
\Yst\ to the \Yst\ values from SPS modelling with a fixed
Milky-Way type IMF, $\Upsilon_{\star,{\rm MW}}$, assumed as a
Chabrier IMF \citep{TRN13_SPIDER_IMF}. We see in Fig.~\ref{fig:FJ}
that at low masses the data are less scattered and MOND with a
Chabrier IMF ($\dimf=1$ line) predicts \sige\ values that agree on
average with the \atlas3d\ observations. At higher masses, the
\sige\ are underpredicted by a factor of $\sim$~1.5 on average,
and require a bottom-heavier IMF for a good match.

Our initial impression from this simple check is that MOND is
discordant with a universal IMF. However, there are additional
correlations with \Re\ to consider which would require a thorough
analysis of the fundamental plane (cf. \citealt{Dutton+13}). We
will instead turn from the space of observables to model space,
where we adjust the input parameters in order to better fit the
data. We also notice that the IMF variation is mild if considered
in terms of stellar mass, while in the following we will discuss
the variable IMF scenario in terms of \sige\
(\citealt{TRN13_SPIDER_IMF}).

\subsection{The acceleration scale}

Our first exercise in model fitting is to consider an alternative
value of the universal constant \a0, thus allowing for relative
systematics between late-type and early-type galaxy modelling. We
treat \a0\ as a free parameter for each of the \atlas3d\ galaxies,
where the goal is to see if the ensemble of \a0\ estimates
scatters around a single consensus value.

\begin{figure}
\centering \psfig{file=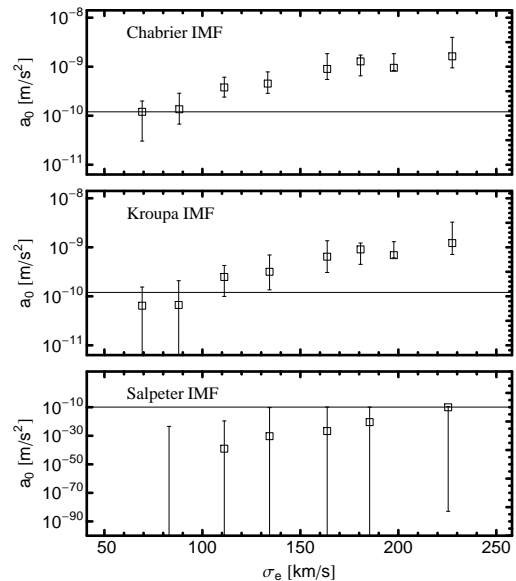,
width=0.4\textwidth}\caption{Inferred MOND acceleration scale vs.\
velocity dispersion. For each \sige-bin, the median and 25--75th
quantiles are shown. From top to bottom we adopt Chabrier, Kroupa
and Salpeter IMF. The standard value for \a0\ is marked with a
horizontal line. MOND is incompatible with any universal
IMF.}\label{fig:a0}
\end{figure}

Fig.~\ref{fig:a0} shows the results, where the galaxies have been
placed in some bins of \sige. Assuming a Chabrier IMF (top panel),
we find that on average, the galaxies are fitted with $\a0 \sim 5
\times 10^{-10}\, \rm m \,s^{-2}$, larger than the standard value
of $1.2 \times 10^{-10}\, \rm m \, s^{-2}$. This is too large a
difference to attribute to errors, and we conclude that MOND
requires more mass in the central parts of ETGs. Smaller \a0\
values are found if we assume a \cite{Kroupa01} IMF (middle
panel). If we instead adopt a Salpeter IMF (bottom panel), we
indeed find that, on average, $a < \a0$ is found. For a large
fraction of galaxies the inferred \a0\ values are very small,
departing from the standard value by several orders of magnitudes
and approaching Newtonian gravity. However, this is not the whole
story, as there is a residual trend for \a0\ to increase with
\sige\ (for both choices of $\mu(x)$). Since
again, \a0\ is meant to be a universal constant, we conclude that
MOND is incompatible with a universal IMF, and we next examine IMF
variations.

\subsection{The variable IMF scenario}

\begin{figure*}
\psfig{file=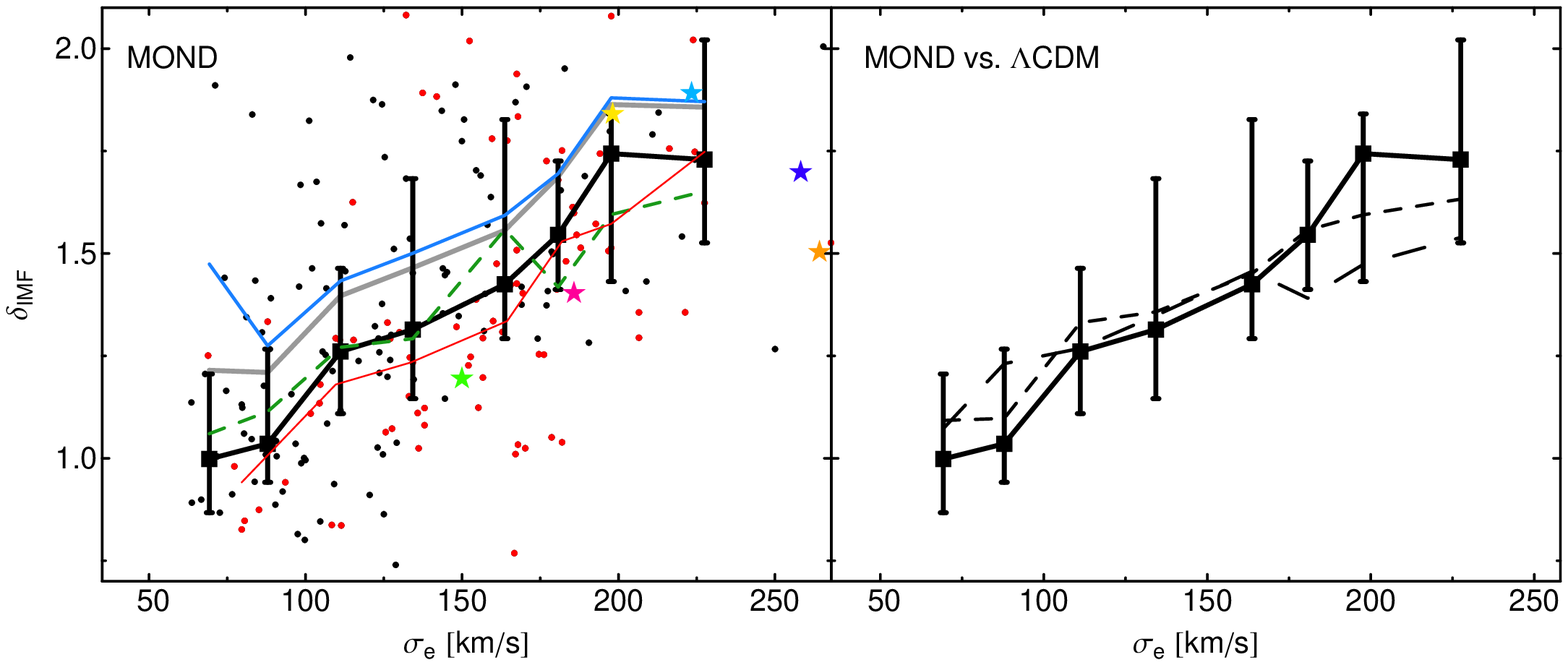, width=0.74\textwidth}
\includegraphics[trim= 27.0mm -10mm 14mm 0mm, width=0.25\textwidth,clip]{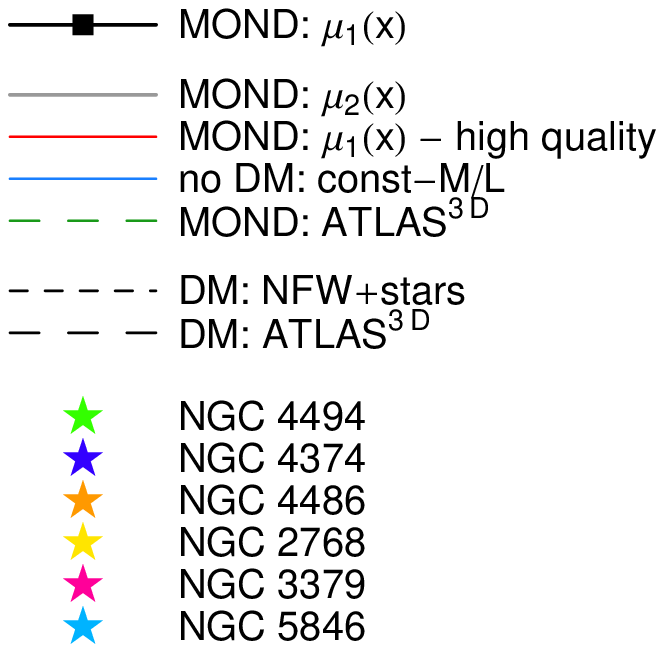}
\caption{IMF mismatch parameter $\delta_{\rm IMF} = \Yst /
\Yst_{\rm MW}$ as a function of velocity dispersion \sige. Thick
black line and squares with bars are for the medians and 25-75th
quantiles for our results adopting the interpolating function
$\mu_{1}(x)$ (left and right panels). Left panel. Single data are
plotted as black points while red ones are for the subsample with
high data quality (\citealt{Cappellari+13_ATLAS3D_XV}). Stars are
for a subsample of relevant galaxies. The red line is the median
trend adopting $\mu_{1}(x)$ and the high-quality data. The gray
line is for the medians assuming the alternative interpolating
function $\mu_{2}(x)$. Blue line is for a constant-\ML\ profile
with no DM within a standard Newtonian scenario. Green dashed line
is obtained using the ``JAM'' dynamical modelling and Eq.
\ref{eq:MOND_JAM}. Right panel. Short-dashed line is for the
medians assuming a standard NFW + S\'ersic model as in
\citet{TRN13_SPIDER_IMF}, while long-dashed one is for medians
using the results from the dynamical analysis in
\citet{Cappellari+13_ATLAS3D_XX}.}\label{fig:IMF}
\end{figure*}

We now return to fixing \a0\ to its standard value, and instead
allow $\Upsilon_*$ to vary on a galaxy-by-galaxy basis in order
for MOND to fit the data. The results are discussed in terms of
the IMF mismatch, \dimf, and plotted as a function of \sige\ in
Fig.~\ref{fig:IMF}. Assuming $\mu_1(x)$, there is a striking
systematic variation from $\dimf \sim 1.0$ (Chabrier-like) for the
lowest-\sige\ galaxies, to $\dimf \sim 1.7$ (near-Salpeter) at the
highest \sige. With an alternative interpolating formula,
$\mu_2(x)$, the MONDian effects are weaker and the implied $\dimf$
values are slightly higher, but the trend with \sige\ remains. The
results are also unchanged if the high quality data are adopted
(see red line in Fig.~\ref{fig:IMF} for $\mu_1(x)$).

Another piece of information is given by comparison with the
results from a constant-\ML\ model with no DM within a Newtonian
scenario (see purple line in the left panel of
Fig.~\ref{fig:IMF}). The only difference with the MOND models is
the change of the gravity theory. This model gives \Yst\ values
which are, on average, $\sim 0.05$ dex larger than the MOND
results using $\mu_{1}(x)$, and quite similar to the ones using
$\mu_{2}(x)$, suggesting that $\mu_{2}(x)$ gives a very tiny
modification of the gravity. We conclude that {\it MOND requires a
strong IMF variation} in order to be consistent with the \atlas3d\
data.

Our dynamical models are limited in their assumptions of
sphericity and isotropy.  We explore radially anisotropic models
with values of $\beta=+0.4$, which produce elevated \sige\ values
but only enough to reduce \dimf\ by $\sim$~10\%. The general
effect of galaxy flattening would be for a spherical model to
over- and under-estimate the mass when a galaxy is edge-on and
face-on, respectively. We have analyzed a relatively face-on
subsample by selecting only the roundest galaxies (ellipticity at
\Re\ of $\epsilon_{\rm e} < 0.2$). The ensuing reduction in \dimf\
is very weak, and does not negate the trend with \sige.

As a final check, we make use of the self-consistent ``JAM''
dynamical modelling results, $\Upsilon_{\rm JAM}$, from
\citet{Cappellari+13_ATLAS3D_XV}, which assume that mass follows
the light and include flattening, anisotropy and more detailed
luminosity profiles. Although these models were constructed using
Newtonian dynamics, we exploit the general insensitivity of the
inferred circular velocities to the details of the mass profile
shapes (e.g., \citealt{Cappellari+13_ATLAS3D_XV}), and use the
results as a fair approximation for what MOND predictions would be
in a fully self-consistent dynamical model. Given a stellar mass
as estimated from SPS, and the associated Newtonian acceleration
$g_N = G \mst / r^{2}$, the corresponding acceleration predicted
by MOND for our default interpolating function is $g =
1/2\,g_N(1+\sqrt{1+4 \a0/g_N})$ (\citealt{Kroupa+10}). After
algebraic manipulation, we find
\begin{equation}
\delta_{\rm IMF} = \frac{\Upsilon_{\rm dyn}}{\Upsilon_{*,\rm MW}}
\left(1+\frac{\a0}{g}\right)^{-1} ,\label{eq:MOND_JAM}
\end{equation}
where $\Upsilon_{\rm dyn}$ is the apparent dynamical \ML\ for an
observer who interprets observations with Newtonian dynamics, and
$\Upsilon_{*,\rm MW}$ is the stellar \ML\ for a fixed Milky-Way
(Chabrier) IMF. Given the standard value for \a0, setting
$\Upsilon_{\rm dyn} \equiv \Upsilon_{\rm JAM}$, $g = G
\Upsilon_{\rm JAM} L(r) / r^{2}$ and calculating all the
quantities at $r=R_{\rm e}$, we estimate $\delta_{\rm IMF}$ on a
galaxy-by-galaxy basis. As shown in the left panel of
Fig.~\ref{fig:IMF}, the results are very similar to ours using
direct, spherical isotropic MOND models. We conclude that the
MONDian IMF variation is robust to the details of the dynamical
models.

\subsection{Comparison to $\Lambda$CDM}

It is now interesting to compare our MOND-based results with those
we obtain within a standard Newtonian scenario. Following
\cite{TRN13_SPIDER_IMF} we adopt an alternative model accounting
for a DM halo. It is based on a \cite{NFW96} profile for the DM
distribution plus a \cite{deVauc48} profile for the stars. For the
virial mass and concentration ($\Mvir$, $\cvir$), we adopt mean
trends for a WMAP5 cosmology \citep{Maccio+08}, while for the
\Mvir--\mst\ relation we used \citet{Moster+10}. Interestingly,
our result for $\mu_{1}$ is fully consistent with the NFW+stars
model and thus with the $\Lambda$CDM expectations (short-dashed
line in the right panel of Fig.~\ref{fig:IMF}). This suggests that
$\Lambda$CDM and MOND are functionally equivalent.

Finally, to illustrate the level of systematic uncertainties for a
method, we have also plotted in the right panel of
Fig.~\ref{fig:IMF} the medians for the DM case (almost similar to
our NFW+stars model) from the results obtained by the Jeans
anisotropic models in \cite{Cappellari+13_ATLAS3D_XX} (see their
Table 1). The agreement is very good.

\section{Conclusions}\label{sec:conclusions}

We have analyzed the dynamical properties of a sample of $\sim
220$ ETGs from the \atlas3d\ survey within a MONDian framework. We
have performed a Jeans analysis of the observed velocity
dispersions and discussed the results in terms of the MOND recipe
details and IMF.

As a preliminary analysis, we have discussed how the observed
\cite{FJ76} relation can be reproduced by MOND, for fixed \a0\ and
IMF (see Fig.~\ref{fig:FJ}). Although not conclusive, we find
hints of non-universality of \a0\ or IMF.

Thus, we determined \a0\ for different choices of the IMF, finding
a trend with \sige\ (Fig.~\ref{fig:a0}), but since \a0\ is meant
to be a universal constant of the theory, we conclude that MOND is
incompatible with a universal IMF. To quantify this result we have
fixed \a0\ to its standard value and allowed \Yst\ to vary.

Following previous literature we focus on the \Yst\ mismatch
relative to a Chabrier IMF, \dimf. Consistently with analysis
involving spectral features (\citealt{Conroy_vanDokkum12b};
\citealt{Ferreras+13}) or dynamical and lensing analysis within a
Newtonian scenario (\citealt{Auger+10}; \citealt{Treu+10};
\citealt{Cappellari+13_ATLAS3D_XX}; \citealt{SPIDER-VI,
TRN13_SPIDER_IMF}) {\it we demonstrate that within a MOND
framework a strong IMF variation is required}
(Fig.~\ref{fig:IMF}). We find a bottom-lighter IMF at low-\sige\
and bottom-heavier at large \sige. Some differences are found in
terms of the interpolating function: $x / (1+x)$ gives \dimf\
values which are fully consistent with $\Lambda$CDM predictions,
while assuming $x / \sqrt{1+x^{2}}$ the gravity is only weakly
modified, such that the \dimf\ values are consistent with what is
found assuming a constant-\ML\ profile with no DM.

Further investigations involving combined dynamical/lensing or
extended kinematical data in ETGs are necessary to probe the
galactic dynamics to their outskirts, where the stellar mass
density is low and the dynamics modification is more important.
Probing different regions of the gravitational potential, we can
provide clearer constraints on the velocity dispersion anisotropy,
the interpolating function and the IMF within a Newtonian scenario
as well as in MOND or different modified gravity theories
(\citealt{Napolitano+12_fR}). Further analysis will investigate a
more general interpolating function, $\mu(x, k_{\rm i})$, and test
whether any combination of $k_{\rm i}$ parameters can remove the
IMF trends. Finally, to have a fully consistent MONDian picture,
one can test whether the varying IMF scenario can ease MOND
tensions in the centers of clusters (\citealt{Angus+10}) and in
gravitational lenses (\citealt{Ferreras+12_MOND_TEVES}) with the
help of a bottom-heavier IMF, and in the very low mass dSph
galaxies in the Local group (\citealt{Kroupa+10}), by the adoption
of a top-heavier IMF.

%%%%%%%%%%%%%%%%%%%%%%%%%%%%%%%%%%%%%%%%%%%%%%%%%%%%%%%%%%%%%%%%%%%%%%%

\section*{Acknowledgments}

We thank the referee for the fruitful comments. We also thank M.
Cappellari and P. Kroupa for helpful discussions. CT was supported
by the Swiss National Science Foundation and the Forschungskredit
at the University of Zurich. CT has received funding from the
European Union Seventh Framework Programme (FP7/2007-2013) under
grant agreement n. 267251.

%%%%%%%%%%%%%%%%%%%%%%%%%%%%%%%%%%%%%%%%%%%%%%%%%%%%%%%%%%%%%%%%%%%%%%%

\bibliographystyle{mn2e}   % (uses file "plain.bst")

%\bibliography{C:/Users/crescenzo/Documents/latex/Bibtex/myrefs}       % expects file "myrefs.bib"

\end{document}